\documentclass[12pt,preprint]{aastex}

\usepackage{epstopdf}

\newcommand\teff{\mbox{$T_\mathrm{eff}$}}

\newcommand\logg{\mbox{$\log g$}}

\newcommand\fmc{\mbox{$f_\mathrm{MC}$}}
\newcommand\kzz{\mbox{$K_\mathrm{zz}$}}
\newcommand\obj{WISEPC J0458$+$64}

\begin{document}

\DeclareGraphicsExtensions{.pdf,.gif,.jpg}

\title{The First Ultra-Cool Brown Dwarf Discovered by the Wide-field Infrared Survey Explorer}
\author{A. Mainzer\altaffilmark{1},Michael C.Cushing\altaffilmark{1},M. Skrutskie\altaffilmark{2},C. R. Gelino\altaffilmark{3},J. Davy Kirkpatrick\altaffilmark{3},T.Jarrett\altaffilmark{3},F. Masci\altaffilmark{3},M. Marley\altaffilmark{4},D. Saumon\altaffilmark{5},E. Wright\altaffilmark{6},R. Beaton\altaffilmark{2},M. Dietrich\altaffilmark{7},P. Eisenhardt\altaffilmark{1},P. Garnavich\altaffilmark{8},O. Kuhn\altaffilmark{9}, D. Leisawitz\altaffilmark{10},K. Marsh\altaffilmark{3}, I. McLean\altaffilmark{6}, D. Padgett\altaffilmark{3},K. Rueff\altaffilmark{8} }

\altaffiltext{(1)}{Jet Propulsion Laboratory, California Institute of Technology, Pasadena, CA 91109 USA}
\altaffiltext{(2)}{University of Virginia, 530 McCormick Road Charlottesville, VA 22904, USA}
\altaffiltext{(3)}{Infrared Processing and Analysis Center, California Institute of Technology, Pasadena, CA 91125, USA}
\altaffiltext{(4)}{NASA Ames Research Center, Mountain View, CA 94043 USA}
\altaffiltext{(5)}{Los Alamos National Laboratory, P.O. Box 1663, Los Alamos, NM 87545 USA}
\altaffiltext{(6)}{UCLA Astronomy, PO Box 91547, Los Angeles, CA 90095-1547 USA}
\altaffiltext{(7)}{Ohio State University Department of Astronomy, 140 W 18th Avenue, Columbus, OH 43210-1173, USA}
\altaffiltext{(8)}{University of Notre Dame, 225 Nieuwland Science Hall, Notre Dame, IN 46556-5670 USA}
\altaffiltext{(9)}{Large Binocular Telescope Observatory, University of Arizona, 933 N. Cherry Ave., Tucson, AZ 85721, USA}
\altaffiltext{(10)}{Goddard Space Flight Center, 8800 Greenbelt Rd., Greenbelt, Md., 20771, USA}

\email{amainzer@jpl.nasa.gov}

\begin{abstract}

We report the discovery of the first new ultra-cool brown dwarf found with the Wide-field Infrared Survey Explorer (WISE).  The object's preliminary designation is WISEPC J045853.90$+$643451.9.  Follow-up spectroscopy with the LUCIFER instrument on the Large Binocular Telescope indicates that it is a very late-type T dwarf with a spectral type approximately equal to T9.  Fits to an IRTF/SpeX 0.8-2.5 $\mu$m spectrum to the model atmospheres of Marley and Saumon indicate an effective temperature of approximately 600 K as well as the presence of vertical mixing in its atmosphere.  The new brown dwarf is easily detected by WISE, with a signal-to-noise ratio of $\sim$36 at 4.6 $\mu$m.  Current estimates place it at a distance of 6 to 10 pc.  This object represents the first in what will likely be hundreds of nearby brown dwarfs found by WISE that will be suitable for follow up observations, including those with the James Webb Space Telescope.  One of the two primary scientific goals of the WISE mission is to find the coolest, closest stars to our Sun; the discovery of this new brown dwarf proves that WISE is capable of fulfilling this objective.

\end{abstract}

\section{Introduction}
Brown dwarfs (BDs) are star-like objects with masses too low to sustain core fusion of hydrogen to helium.  Although first predicted to exist in the early 1960's \citep{Kumar,Hayashi}, BDs were not discovered in bulk until decades later by deep, large-area surveys such as the Two Micron All Sky Survey \citep[2MASS;][]{Skrutskie}, the Sloan Digital Sky Survey \citep[SDSS;][]{York}, and the DEep Near-Infrared Southern Sky Survey \citep[DENIS;][]{Epchtein}. Strong methane absorption characterizes the coolest observed class of brown dwarfs, providing the hallmark of the ``T'' spectral type.  To date, more than 200 T dwarfs have been discovered (DwarfArchives.org). T dwarfs have effective temperatures ranging from $\sim$1300 K down to $\sim$450 K \citep{Lucas,Burgasser,Burningham08,Delorme,Golimowski,Warren}, and the coldest T dwarfs are characterized by deep absorption bands of CH$_{4}$, H$_{2}$O, H$_{2}$ and NH$_{3}$ appearing throughout the near- and mid-infrared portions of their spectrum \citep{Borysow,Irwin,Leggett07}.  No census of our solar neighborhood can be complete without discovering the lowest-temperature brown dwarfs, as it is anticipated that they exist in numbers rivaling those of normal stars \citep{Burningham10,Lodieu,Moraux,Reid}.  Additionally, the spatial density of ultra-cool and low mass BDs can be used as a sensitive probe of the lowest masses that can be produced via different star formation processes \citep{Bate, Burgasser04}.  

One of the two primary science objectives for the Wide-field Infrared Survey Explorer (WISE) is to find these coldest brown dwarfs, which represent the final link between the lowest mass stars and the giant planets in our own solar system.  WISE is a NASA Medium-class Explorer mission designed to survey the entire sky in four infrared wavelengths, 3.4, 4.6, 12 and 22 $\mu$m \citep{Wright,Liu,Mainzer}.  WISE consists of a 40 cm telescope that images all four bands simultaneously every 11 seconds. It covers nearly every part of the sky a minimum of eight times, ensuring high source reliability, with more coverage at the ecliptic poles.  Astrometric errors are less than 0.5 arcsec with respect to 2MASS \citep{Wright}.  The preliminary estimated SNR$=$5 point source sensitivity on the ecliptic is 0.08, 0.1, 0.8 and 5 mJy in the four bands (assuming eight exposures per band; Wright et al. 2010). Sensitivity improves away from the ecliptic due to denser coverage and lower zodiacal background.  

WISE's two shortest wavelength bands, centered at 3.4 and 4.6 $\mu$m ($W1$ and $W2$, respectively), were specifically designed to optimize sensitivity to the coolest types of brown dwarfs \citep{Kirkpatrick10}.  In particular, BDs cooler than $\sim$1500 K exhibit strong absorption due to the $\nu_{3}$ band of CH$_{4}$ centered at 3.3 $\mu$m, with the onset of methane absorption at this wavelength beginning at $\sim$1700 K \citep{Noll}.  The as-measured sensitivities of 0.08 and 0.1 mJy in $W1$ and $W2$ allow WISE to detect a 300 K BD out to a distance of $\sim$8 pc, according to models by \citet{Marley} and \citet{Saumon}. 

\section{Observations of Known L and T Dwarfs with WISE}
In order to facilitate the search for new BDs in the WISE dataset, we first examined the WISE fluxes and colors of known BDs. The $W1-W2$ and IRAC $[3.6]-[4.5]$ colors of a sample of L and T dwarfs are shown in Figure 1a along with their Spitzer/IRAC colors \citep{Patten}. The WISE $W1$ band spans the deep H$_{2}$O and CH$_{4}$ absorption features found in the coolest BDs, while the $W2$ band is centered on the window near 5 $\mu$m where much of the flux of these objects is emitted; it also includes the 4.6 $\mu$m CO absorption feature and the 4.2 $\mu$m CO$_{2}$ feature \citep{Yamamura}. While the $W1$ and $W2$ bands are superficially similar to the Spitzer/IRAC bands 1 and 2 \citep[which were also designed to isolate cool BDs;][]{Fazio}, the WISE $W1$ band is wider to improve discrimination between normal stars and ultra-cool BDs (Wright et al. in preparation). This results in a systematically larger $W1-W2$ color compared to [3.6]$-$[4.5].  Figures 1b and 1c show that $J-W2$ and $H-W2$ increase with spectral type, with $H-W2$ showing the smoothest correlation between color and spectral type for BDs later than T5, in good agreement with \citet{Leggett10}; these indices are likely to be a valuable means of identifying new L and T dwarfs. Table 1 contains the WISE photometry of the small sample of L and T dwarfs shown in the figures, and Figure \ref{fig:RSR} shows a model of a cool BD along with the WISE $W1$ and $W2$ transmission curves.  

\section{Search for New Brown Dwarfs with WISE}
Several methods were employed for searching for new BD candidates in the WISE dataset.  The WISE preliminary data processing pipeline produces calibrated frames for each of the individual exposures.  These images have been dark-subtracted, flat-fielded, distortion corrected, have had artifacts flagged and sources extracted.  Once the individual images are produced, they are coadded to produce a single frame for each inertial position.  

Searches for new BD candidates were performed on both the single-frame and multi-frame coadded mosaic source lists. The coadd search provided the maximum depth and sensitivity since the coadded frames generally feature the combination of at least eight frames per position.  The coadd search was performed by searching the source lists for objects with a $W1-W2 \geq2.0$, a $W2$ signal-to-noise ratio (SNR) $\geq$10, although sources were not required to be detected in $W1$ (limiting magnitudes are given to 2.5 $\sigma$).  Sources were required to have been observed a minimum of four times in order to ensure reliability. Sources affected by artifacts such as diffraction spikes and latent images were excluded, and the reduced chi-squared point spread function goodness-of-fit parameter was required to be consistent with a point source (w2rchi2$\leq$3, where the quantity w2rchi2 represents the goodness of fit of a point source model to the set of observed pixel values in band $W2$, whereby a value close to unity indicates consistency).  No requirement on the $W1$ signal-to-noise ratio was levied in order to increase sensitivity to the coolest BDs, which can fall below the $W1$ sensitivity limits.  Subsequent searches were performed with varied criteria, including requiring the $W4$ SNR to be $\leq$5, and only sources with high reliability were accepted (as determined by comparing the number of detections to the number of coverages).  Once sources meeting these criteria were extracted from the WISE data, they were compared with images from the Digitized Sky Survey (DSS), 2MASS, and SDSS.  BD candidates were required to have blue and red optical fluxes from DSS and SDSS (where SDSS data were available) consistent with expectations for very late T dwarfs observed with WISE.  Cold WISE BD candidates were required to be unseen in DSS or SDSS, or in some cases visible at the longest wavelengths only in these surveys.   

A source, given the provisional designation WISEPC J045853.90$+$643451.9 (hereafter \obj), was located in the coadd source list that met all of the above criteria; Table 2 summarizes the astrometry, and Table 3 summarizes the WISE magnitudes and colors for the source.  Although \obj\ is faintly visible in the 2MASS $J$ band original scans, slightly offset from the WISE-derived position, the source was not bright enough to have been included in either the 2MASS All-Sky Point Source Catalog or Reject Table.  Aperture photometry of the $J$-band flux using standard IDL astronomy library routines reveals it to be detected at 3 $\sigma$.  Figure \ref{fig:0458_finder} shows the WISE, 2MASS and DSS finder images.  At the position of the BD candidate, there are 13 separate images from WISE.  The signal-to-noise ratios in the resulting coadded data in $W1$ are 11.9 and 36.3 in $W2$.  The source was not detected in $W3$ due to contaminating cirrus, which can confound photometry at these wavelengths.  The object is not detected in $W4$ to a limiting magnitude of 9.3 (Vega system), or 1.70 mJy. The images of \obj\ were searched for widely separated companions, but none were found.  No known nearby stars with common proper motion were found.  

\section{Follow-Up Observations of \obj}
\subsection{Fan Mountain - FanCam}
$YJH$ photometry of \obj\ was obtained on 2010 March 17 (UT) with FanCam, a HAWAII-1 based infrared imager operating at the University of Virginia's Fan Mountain 31" telescope \citep{Kanneganti}.  The source position was dithered by approximately 10" between 30 s ($JH$) or 60 s ($Y$) exposures during total exposures of 40, 35, and 60 minutes at $Y$, $J$ and $H$-band respectively.  The FanCam field of view (FOV) is 8.7' x 8.7' (0.51"/pixel).  The combined, dithered exposures covered 7' x 7', providing several 2MASS stars for photometric reference within the exposure area.  A median ``sky'' frame was subtracted from each individual exposure prior to flat fielding with the median background level subsequently restored to the image. 

$YJH$ aperture photometry was computed for \obj\ using an aperture with a radius of 3 pixels.  The zero points for the $JH$-bands were computed using stars in the FOV with measured 2MASS magnitudes since the $JH$-band filters in FanCam are based on the 2MASS system.   In order to derive the $Y$-band zero point, we first computed the $Y$-band magnitudes of stars in the FOV using their 2MASS $J$ and $K_{s}$ magnitudes and the transformation given by \citet{Hamuy}.  The final uncertainty in the magnitudes include the photon noise from the sky and source, the read noise, and the uncertainty in the zero point.  The resulting magnitudes and uncertainties are given in Table 3.  

\subsection{Large Binocular Telescope - LUCIFER}
The LBT Near Infrared Spectroscopic Utility with Camera and Integral Field Unit for Extragalactic Research (LUCIFER) near-infrared camera/spectrograph \citep{Mandel}, operating in long-slit grating spectroscopy mode, provided 1.5-2.3 $\mu$m spectra of \obj\ at the Gregorian focus of the Large Binocular Telescope's SX (``left'') 8.4-meter primary mirror on 2010 March 19 (UT).  The 0.75" slit spanned 3.0 pixels and provided a spectral resolving power $R$ ($\equiv \lambda / \delta \lambda $) ranging from 1300 at 1.6 $\mu$m to 1700 at 2.2 $\mu$m dispersing the light in a single order across the array.  The orientation of the 200 lines/mm ``H+K'' grating centered 1.98 $\mu$m on the array.  Observations consisted of 300 second exposures with the source nodded between two slit positions separated by 20 arcseconds in declination following an ABBA position sequence. Three repetitions of this sequence provided for 60 minutes of on-source integration.  A0 telluric standards were observed both before and after the observation.

The spectra were processed using a modified version of the Spextool data reduction package \citep{Cushing}. Each pair of exposures was subtracted and then flat fielded. The spectra were then extracted and wavelength calibrated using sky emission features. The raw spectrum of \obj\ was corrected for telluric absorption using the A0 V star HD 33541 and the technique described by \citet{Vacca}.  The spectrum was rebinned to have 1 pixel per resolution element resulting in a final signal-to-noise ratio of 20 and 10 in the $H$ and $K$ bands respectively. 

\subsection{NASA Infrared Telescope Facility - SpeX}
We obtained a 0.8$-$2.5 $\mu$m spectrom of \obj\ on 2010 Sep 12 (UT) using SpeX \citep{Rayner2003} on the 3.0 m NASA Infrared Telescope Facility on Mauna Kea.  We used the low-resolution prism mode with the 0".5 wide slit which yielded an average resolving power of $R=$150. A series of 120 sec exposures at two positions along the 15" slit were obtained to facilitate sky subtraction.  Observations of the A0 star HD 14632 were obtained after the science exposures for telluric correction and flux calibration purposes as well as flat and arc lamp exposures. 

The data were reduced using Spextool \citep{Cushing}, the IDL-based data reduction package for SpeX.   Given the faintness of \obj, the images were first pair-subtracted and then averaged together robustly in order to facilitate extraction.  The spectra were then optimally extracted \citep[e.g.][]{Horne} and wavelength calibrated using argon lamp exposures.  The raw spectrum was then corrected for telluric absorption and flux-calibrated using the observations of the A0 V star HD14632 and the technique described in \citep{Vacca}.  The SNR of the final spectrum ranges from a maximum of $\sim$40 in the $J$ band to $\sim$5 in the $K$ band.  The spectrum was absolutely flux calibrated as described in \citet[see Equation 1]{Rayner} using the FanCam $J$- and $H$-band magnitudes. The agreement between the LUCIFER and SpeX data is excellent (Figure \ref{fig:spex_lucifer}).

\section{Analysis}
\subsection{Spectral Analysis}
The LUCIFER and SpeX spectra are shown in Figure \ref{fig:spectral_comparison}.  The spectra exhibit the deep absorption features of CH$_4$ and H$_2$O indicative of a very late T dwarf.  The spectrum of \obj\ can be compared to the NIR spectra of the very late T dwarfs 2MASS J0415$-$09 \citep{Burgasser02}, ULAS J003402.77$-$005206.7 \citep[hereafter ULAS J0034$-$00;][]{Warren} , CFBDS J005910.90$-$011401.3 \citep[CFBDS J0059$-$00;][]{Delorme}, ULAS J133553.45$+$113005.2 \citep[ULAS J1335$+$11;][]{Burningham08}, and ULAS J141623.94$+$134836.3 \citep[ULAS J1416$+$13;][]{Scholz}.  As shown in Figure \ref{fig:spectral_comparison}, \obj\ appears to best match the spectrum of the T9 dwarfs, although formal assignment of a spectral type will have to wait for the identification of spectral standards for objects this late. 

The atmospheric parameters of \obj\ were estimated by comparing the SpeX spectrum and the WISE photometry to the model spectra of \citet{Marley} and \citet{Saumon}. We used a grid of 80 cloudless models with \teff\ ranging from 500 to 1000 K in steps of 100 K, \logg\ = 4.0, 4.5, 5.0, 5.5 [cm s$^{-2}$], [M/H] = 0, $-$0.3, and \kzz\ = 0, $10^4$ cm$^{2}$ s$^{-1}$. \kzz\ is the eddy diffusion coefficient and ranges from $\sim$10$^2$ cm$^{2}$ s$^{-1}$ to 10$^5$ cm$^2$ s$^{-1}$ in planetary atmospheres \citep[e.g.,][]{Saumon06}. \kzz\ $=$0 corresponds to an atmosphere without vertical mixing in the radiative zone where chemical equilibrium prevails.  The models were first smoothed to the resolution of the spectra and interpolated onto the wavelength scale of the data. In order to include the WISE photometry in the comparison, we converted the observed W1 and W2 magnitudes to average fluxes using the following zero points on the Vega system: 306.681, 170.663, 29.0448 \& 8.2839 Jy in bands $W1$ through $W4$ and using the relative spectral response curves from \citep{Wright, Jarrett}.

We identified the best fitting model spectra in the grid using the goodness-of-fit statistic $G_k$ ($w_i$=1; all data points receiving equal weight) described in \citet{Cushing08}. For each model $k$, we compute the scale factor $C_k$=$(R/d)^2$, where $R$ and $d$ are the radius and distance of the dwarf respectively, that minimizes $G_k$. The best-fitting model is identified as having the global minimum $G_k$ value. To access the uncertainty in the selection process, we run a Monte Carlo simulation using both the uncertainties in the individual spectral and photometric points and the overall absolute flux calibration of the SpeX spectrum \citep[see][]{Cushing08, Bowler99}. We report the quality of the fit between the data and model $k$ as \fmc, the fraction of Monte Carlo simulations in which model $k$ is identified as the best fitting model. 

The best-fitting model has \teff=600 K, \logg=5.0, [Fe/H]=0, \kzz=10$^4$ cm$^2$ s$^{-1}$, and \fmc=0.66 while the second best-fitting model has \teff=600 K, \logg=5.5, [Fe/H]=0, \kzz=10$^4$ cm$^2$ s$^{-1}$, and \fmc=0.34. Given their \fmc values, these are the only two models in the grid that fit the data given the uncertainties. Figure \ref{fig:ModComp} shows the SpeX spectrum and WISE photometry of \obj\ along with the best-fitting model.  The spectrum is well-matched by the model, but the $W2$ point is discrepant by over 10 $\sigma$.  Although Figure \ref{fig:spectral_comparison} shows that \obj\ appears to show suppressed K band flux relative to other late T dwarfs, the interpretation of the feature is uncertain as the models for objects in this temperature regime are as yet not well-validated.  Suppressed $K$-band flux has previously been ascribed to low metallicity and/or high surface gravity \citep{Leggett09,Stephens}; however, our model fits do not indicate unusual metallicity.  The atmospheric parameters derived herein for \obj\ are consistent with those derived for other late-type T dwarfs \citep[e.g.,][]{Leggett09, Burningham09}.  In particular there is evidence for the presence of vertical mixing in the atmosphere of \obj\ as \kzz=10$^4$ cm$^2$ s$^{-1}$.  However the poor mismatch between the data and the model at 4.6 $\mu$m, a spectral region particularly sensitive to vertical mixing \citep{Leggett07b, Geballe}, makes drawing any conclusions difficult.    

Following \citet{Bowler99}, we can also estimate a \textit{spectrophometric} distance to \obj\ using the scale factor $C_k$. We first estimate the radius of \obj\ using the derived (\teff, \logg) values and the evolutionary models of \citet{Saumon}. Using errors given by half the grid spacing of 50 K and 0.25 dex, we estimate a distance of 6$-$8 pc for \obj.ú This distance is in rough agreement with the photometric distance of 9.0$\pm$1.9 pc, using the apparent $J$ magnitude of 17.47$\pm$0.05 and absolute magnitude M$_J$ = 17.7$\pm$0.45. This was computed from a weighted average of M$_J$ for the late T dwarfs Wolf 940B \citep{Burningham09}, ULAS0034$-$00 \citep{Smart}, and ULAS141623.94$+$134836.3 \citep{Scholz}, using the worst-case error.    

\subsection{Astrometry}
The source is not present in either the 2MASS All-Sky Point Source Catalog or the Reject Catalog, but it is faintly visible in the 2MASS $J$ image as shown in \ref{fig:0458_finder}. We performed a Gaussian fit to the source in the 2MASS $J$ image to measure its position. Astrometry at three epochs (2MASS, WISE, FanCam)  is listed in Table 2.  A linear least-squares fit to the measured positions, weighted by their astrometric errors, gives a motion of 0.1968$\pm$0.0291"/yr in right ascension and 0.1593$\pm$0.0291"/yr in declination over 11.2 years, or a total motion of 0.256$\pm$0.032"/yr.  

\section{Discussion}
One of the key questions in brown dwarf science is whether or not the spectral sequence ends with the T dwarfs, or whether another spectral type (with a proposed designation of Y; \citet[]{Kirkpatrick99}) is justified.  Currently, only about two dozen objects with spectral types later than T7 are known, and four objects are known to have type T9 or later: ULAS J0034$-$00 \citep{Warren}, CFBDS J0059$-$00 \citep{Delorme}, ULAS J1335$+$11 \citep{Burningham08}, and UGPS J0722$-$05 \citep{Lucas}.  New spectral indices for typing these late-type T dwarfs have been developed as the CH$_4$ absorption band depths may not continue to increase in the NIR with temperatures lower than those of T8/T9 dwarfs \citep{Burningham10}.  We compute the H$_{2}$O$-J$, CH$_{4}-J$, $W_{J}$, H$_{2}$O$-H$, CH${_4}-H$ and CH${_4}-K$ indices given by \citet{Burgasser}, \citet{Burningham08} and \citet{Warren} for \obj\ (see Table 3).  However, as noted in \citet{Burningham08}, these indices must viewed with caution for extremely late T dwarfs; the values we compute for \obj\ tell us only that it is later than T7/T8.  Given the paucity of very late T dwarfs, assignment of a definitive spectral type will await the identification and definition of spectral standards.  As this remarkably cool object was found easily in some of the first WISE data, WISE is likely to find many more similar objects, as well as cooler ones, inevitably producing abundant candidates for the elusive Y class brown dwarfs. Scaling from the Spitzer sample of 4.5 $\mu$m selected ultra-cool BD candidates \citep{Eisenhardt}, we expect to find hundreds of new ultra-cool brown dwarfs with WISE.  We compute that for most likely initial mass functions, WISE has a better than 50\% chance of detecting a cool BD which may actually be closer to our Sun than Proxima Centauri \citep{Wright}.  

\section{Conclusions}
We report the discovery of the first ultra-cool brown dwarf with WISE, an object which is consistent with an extremely late T dwarf.  The object is easily detected by WISE, with signal-to-noise ratio of $\sim$40 at 4.6 $\mu$m. Further, \obj\ lies in projection only 13 degrees off the Galactic midplane.  WISE can pick out ultra-cool brown dwarfs like this one easily even in regions that are traditionally confused, as the number of sources per square  at 4.6 $\mu$m is significantly lower than in optical and near-IR wavelengths.  The WISE survey offers an all-sky survey with extremely uniform and stable instrumental characteristics and survey cadence.  By surveying the entire sky, including the galactic plane, WISE can provide an accurate census of cold BDs in the solar neighborhood, which is necessary for determining the initial mass function at low masses.   \obj, along with UGPS J0722$-$05 and the few other brown dwarfs like it, represents the first in what is anticipated to be a significant population of objects revealed by the WISE survey.  We expect that brown dwarfs discovered by WISE will at last close the remaining gap in temperature between the coolest objects currently known ($\sim$500 K) and potentially those of the giant planets in our own solar system \citep{Wright}.  

\section{Acknowledgments}

\acknowledgments{This publication makes use of data products from the Wide-field Infrared Survey Explorer, which is a joint project of the University of California, Los Angeles, and the Jet Propulsion Laboratory/California Institute of Technology, funded by the National Aeronautics and Space Administration.  We are deeply grateful for the outstanding contributions of all the members of the WISE team.  We thank Roger Griffith for assistance with Figure 2 and Beth Fabinsky for assistance with early searches.  Support for the modeling work of D. S. was provided by NASA through the Spitzer Science Center.  M.C. was supported by an appointment to the NASA Postdoctoral Program at the Jet Propulsion Laboratory, administered by Oak Ridge Associated Universities through a contract with NASA.This publication makes use of data products from the Two Micron All Sky Survey (2MASS).  2MASS is a joint project of the University of Massachusetts and the Infrared Processing and Analysis Center/California Institute of Technology, funded by the National Aeronautics and Space Administration and the National Science Foundation. This research has made use of the NASA/IPAC Infrared Science Archive (IRSA), which is operated by the Jet Propulsion Laboratory, California Institute of Technology, under contract with the National Aeronautics and Space Administration. Our research has benefitted from the M, L, and T dwarf compendium housed at DwarfArchives.org whose server was funded by a NASA Small Research Grant, administered by the American Astronomical Society. We are also indebted to the SIMBAD database, operated at CDS, Strasbourg, France.  The Digitized Sky Surveys were produced at the Space Telescope Science Institute under U.S. Government grant NAG W-2166. The images of these surveys are based on photographic data obtained using the Oschin Schmidt Telescope on Palomar Mountain and the UK Schmidt Telescope.  The Second Palomar Observatory Sky Survey (POSS-II) was made by the California Institute of Technology with funds from the National Science Foundation, the National Geographic Society, the Sloan Foundation, the Samuel Oschin Foundation, and the Eastman Kodak Corporation. The Oschin Schmidt Telescope is operated by the California Institute of Technology and Palomar Observatory.  We thank Richard Green and the LBT staff for making the LUCIFER observations possible.  The LBT is an international collaboration among institutions in the  United States, Italy and Germany. LBT Corporation partners are: The  University of Arizona on behalf of the Arizona university system;  Istituto Nazionale di Astrofisica, Italy; LBT Beteiligungsgesellschaft,  Germany, representing the Max-Planck Society, the Astrophysical  Institute Potsdam, and Heidelberg University; The Ohio State University,  and The Research Corporation, on behalf of The University of Notre Dame,  University of Minnesota and University of Virginia.  The LUCIFER Project is funded by the Bundesministerium f\"{u}r Bildung und  Forschung (BMBF). It is a collaboration of five German institutes,  Landessternwarte Heidelberg, Max  Planck Institut f\"{u}r Astronomie (Heidelberg), Max  Planck Institut f\"{u}r Extraterrestrische Physik (Garching), Fachhochschule f\"{u}r Technik und Gestaltung  (Mannheim), Astronomisches Institut der  Universit\"{a}t Bochum.
}

\clearpage

\begin{deluxetable}{ccccccc}
\tabletypesize{\small}
\tablecolumns{7}
\tablecaption{Preliminary Photometry of A Sample of Known Brown Dwarfs Observed with WISE.  Spectral types for L dwarfs are based on optical spectroscopy; T dwarfs are based on infrared spectroscopy.}
\tablehead{
\colhead{Object} & \colhead{Spectral Type} & \colhead{$W1$} & \colhead{$W1$ Error} & \colhead{$W2$} & \colhead{$W2$ Error} & \colhead{$W1-W2$} }
\startdata
2MASS J04510093$-$3402150\tablenotemark{a} & L0.5 & 11.933 & 0.016 & 11.690 & 0.015 & 0.243\\
2MASS J08354256$-$0819237\tablenotemark{a} & L5 & 10.403 & 0.025 & 10.04 & 0.020 & 0.363\\
2MASS J16322911$+$1904407\tablenotemark{b} & L8 & 13.151 & 0.029 & 12.637 & 0.024 & 0.514\\
SDSSp J083717.22$-$000018.3\tablenotemark{c} & T0$-$T1 & 15.512 & 0.056 & 14.697 & 0.079 & 0.815\\
IPMS J013656.57$+$093347.3\tablenotemark{d} & T2.5 & 11.922 & 0.026 & 10.943 & 0.024 & 0.979\\
2MASSI J1546291$-$332511\tablenotemark{e} & T5.5 & 15.044 & 0.072 & 13.447 & 0.037 & 1.597\\
SDSSp J162414.37$+$002915.6\tablenotemark{f} & T6 & 15.2 & 0.053 & 13.146 & 0.030 & 2.054\\
Gliese 570D\tablenotemark{g} & T7 & 15.029 & 0.119 & 12.151 & 0.045 & 2.878\\
2MASSI J0415195$-$093506\tablenotemark{e} & T8 & 15.207 & 0.051 & 12.28 & 0.022 & 2.927\\
UGPS J072227.51$-$054031.2\tablenotemark{h} & T9+ & 15.168 & 0.047 & 12.197 & 0.027 & 2.971\\
\enddata
\tablenotetext{a}{\citet{Cruz}}
\tablenotetext{b}{\citet{Kirkpatrick99}}
\tablenotetext{c}{\citet{Leggett00}}
\tablenotetext{d}{\citet{Artigau}}
\tablenotetext{e}{\citet{Burgasser02}}
\tablenotetext{f}{\citet{Strauss}}
\tablenotetext{g}{\citet{Burgasser00}}
\tablenotetext{h}{\citet{Lucas}}

\end{deluxetable}

\begin{deluxetable}{cccccc}
\tabletypesize{\small}
\tablecolumns{6}
\tablecaption{Astrometry of \obj.}
\tablehead{
\colhead{RA (J2000, $^{\circ}$)} & \colhead{Dec (J2000, $^{\circ}$)} & \colhead{Ra err ($"$)} & \colhead{Dec err ($"$)} & \colhead{MJD} & \colhead{Reference} }
\startdata
74.7245712 & 64.5810776 & 0Ó.1696 & 0Ó.1703 & 55262.18273\tablenotemark{a} & WISE Pass 1 Database \\
74.723142 & 64.580773 & 0Ó.3 & 0Ó.3 & 51179.32788 & 2MASS Images\\
74.724583 & 64.581224 & 0Ó.2 & 0Ó.2 & 55272.03133 & FanCam images\\
\enddata

\tablenotetext{a}{This is the average of the date of the earliest frame (MJD= 55261.554191) and that of the latest frame (MJD=55262.811271) contributing to the coadded image.}

\end{deluxetable}

\begin{deluxetable}{cc}
\tabletypesize{\small}
\tablecolumns{2}
\tablecaption{Photometry, Proper Motion, and Spectral Indices of \obj. }
\tablehead{
\colhead{Parameter Name} & \colhead{Value (mag)}}
\startdata
$J$ (2MASS) & 16.8$\pm$0.3 \\
$H$ (2MASS) & (undetected)\\
$K$ (2MASS) & (undetected)\\
$Y$ (FanCam) & 18.34$\pm$0.07\\
$J$ (FanCam) & 17.47$\pm$0.07\\
$H$ (FanCam) & 17.41$\pm$0.11\\
$W1$ & 16.401$\pm$0.100\\
$W2$ & 13.020$\pm$0.031\\
$W3$ & (undetected - cirrus) \\
$W4$ & ( 9.322 - undetected; lower limit)\\
$Y{_{FC}}-J{_{FC}}$ & 0.87$\pm$0.10\\
$J{_{FC}}-H{_{FC}}$ & 0.06$\pm$0.13\\
$H{_{FC}}-W2$ & 4.39$\pm$0.11\\
$W1-W2$ & 3.381$\pm$0.105\\
Proper motion (arcsec/yr) & 0.256$\pm$0.032\\
Position angle & 51$\pm$7 degrees east of north\\
H$_{2}$O$-J$ & 0.03\\
CH$_{4}-J$ & 0.32\\
$W_{J}$ & 0.46\\
H$_{2}$O$-H$ & 0.10\\
CH$_{4}-H$ & 0.09\\
CH$_{4}-K$ & 0.19\\
\enddata
\end{deluxetable}

\clearpage

\begin{figure}
\figurenum{1a,b,c}
\includegraphics[width=3in]{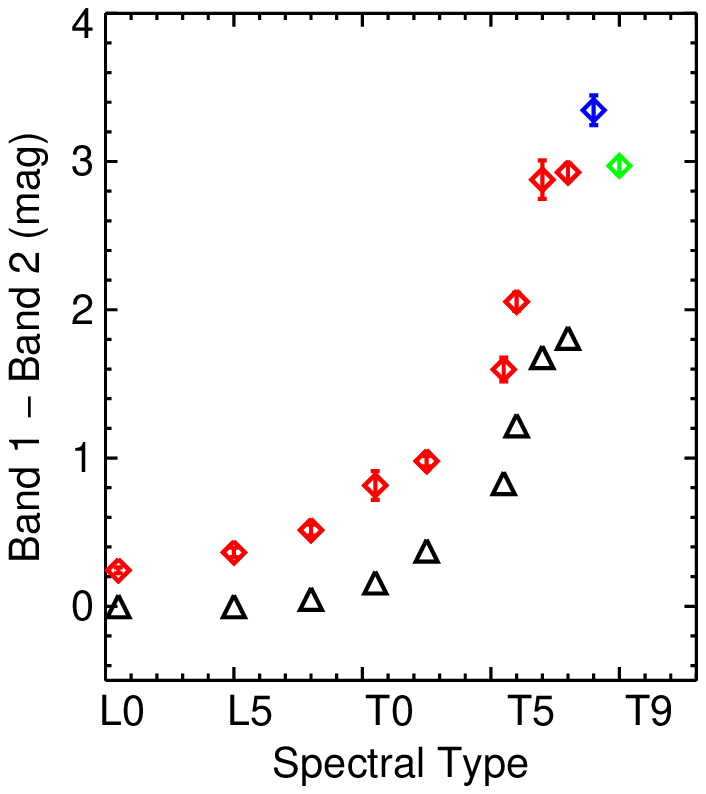}
\plottwo{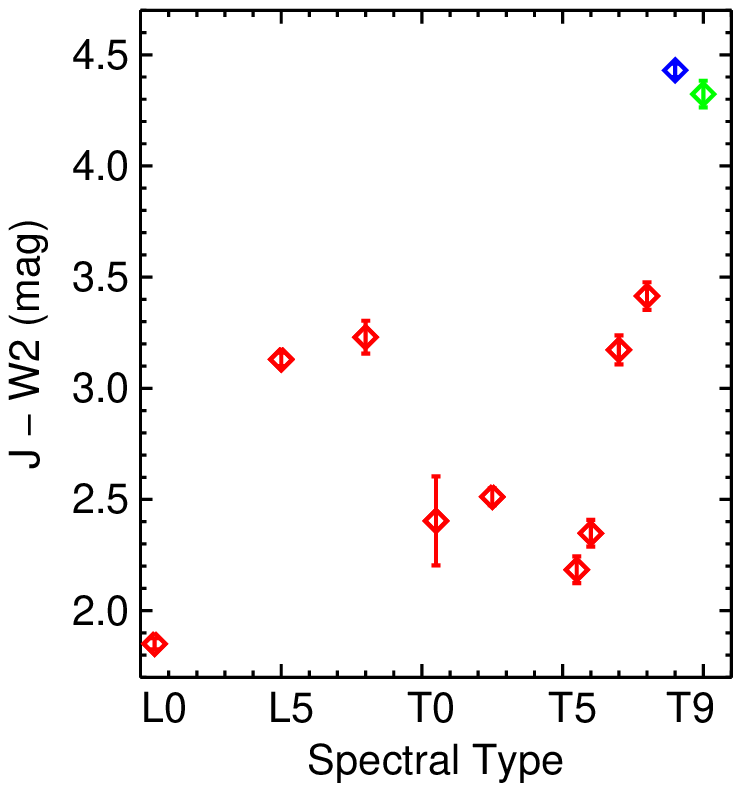}{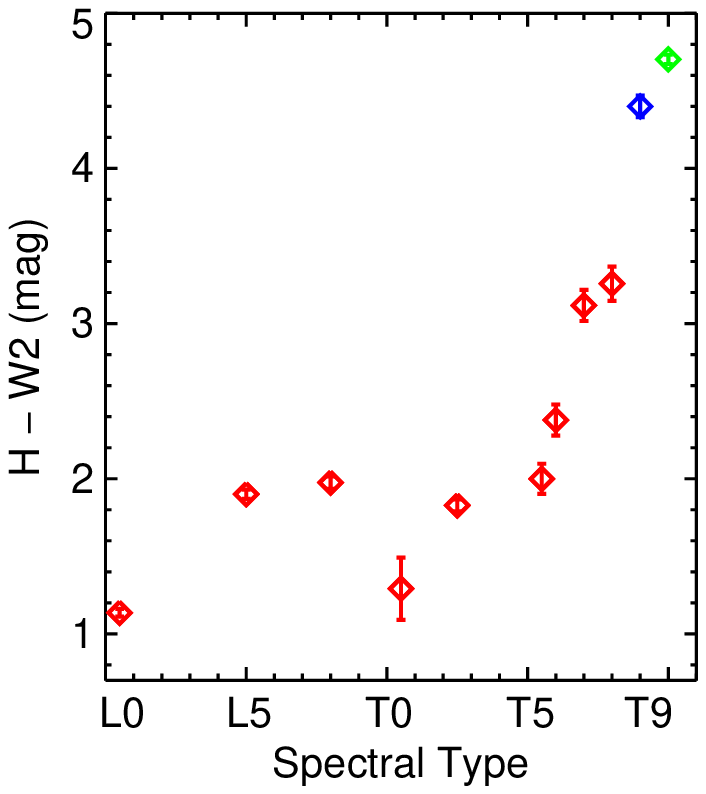}
\caption{\label{fig:W1W2}The WISE $W1-W2$ colors are systematically redder than the IRAC Ch1$-$Ch2 colors for low temperature objects.  WISE measurements are shown as red diamonds; Spitzer/IRAC measurements for the same objects are shown as black triangles \citep{Patten}. \obj\ is shown as a blue diamond, and UGPS J0722$-$05 \citep{Lucas} is shown as a green diamond. \obj\ appears $\sim$0.4 magnitudes redder than UGPS J0722$-$05.  UGPS J0722$-$05 is detected with a measured SNR=39.9 by WISE in the $W2$ image formed from coadding 10 frames.  The $J-W2$ and $H-W2$ colors increase rapidly with spectral type.  $J-W2$ and $H-W2$ are likely to be effective indices for detecting and characterizing late L and early T dwarfs as well as ultra-cool BDs.} 
\end{figure}

\begin{figure}
\figurenum{2}
\plotone{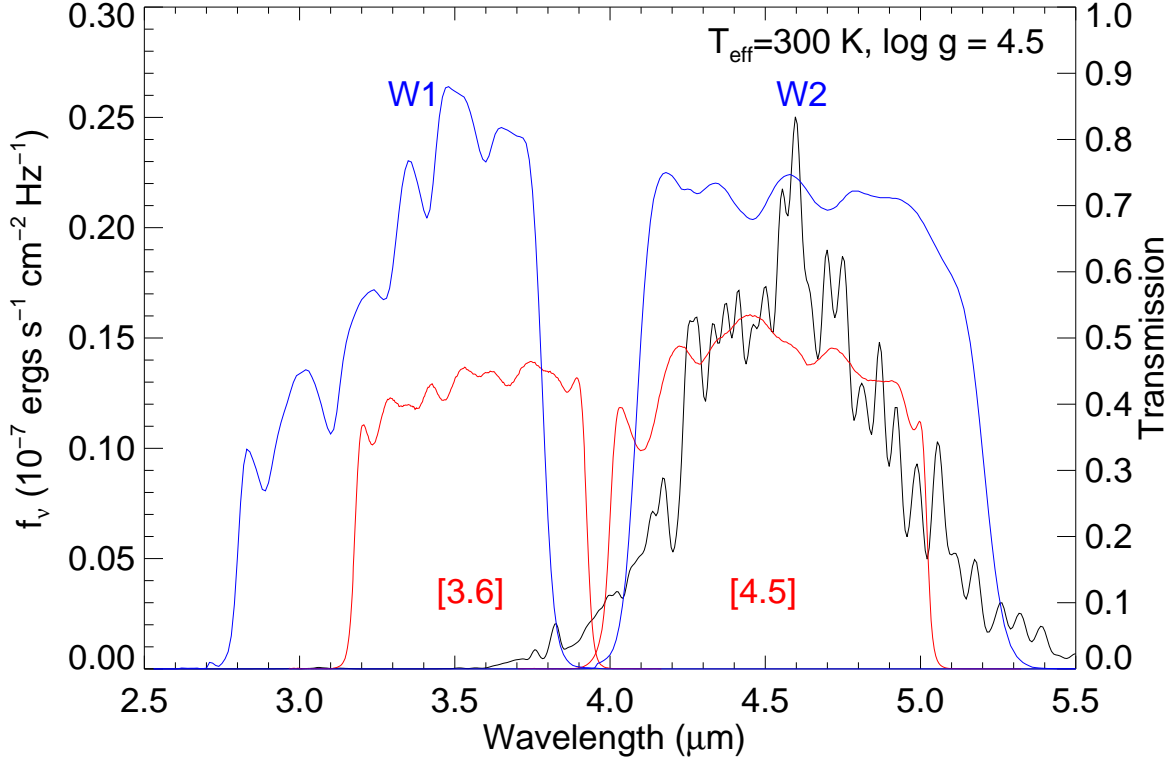}
\caption{\label{fig:RSR}A cloudless model spectrum with \teff=300 K and \logg = 4.5 (cm s$^{-2}$) (M. Marley \& D. Saumon, private communication, 2010) is shown in black along with the relative response functions of the 3.6 and 4.5 $\mu$m bands of IRAC (red) and the W1 and W2 bands of WISE (green).  The positions and widths of the $W1$ and $W2$ bandpasses were specifically designed to sample the strong absorption at $\lambda \sim$3.5 $\mu$m due to CH$_{4}$ and H$_{2}$O and the region relatively free of opacity at $\sim$4.7 $\mu$m in the emergent spectra of cold BDs.}
\end{figure}

\begin{figure}
\figurenum{3}
\plotone{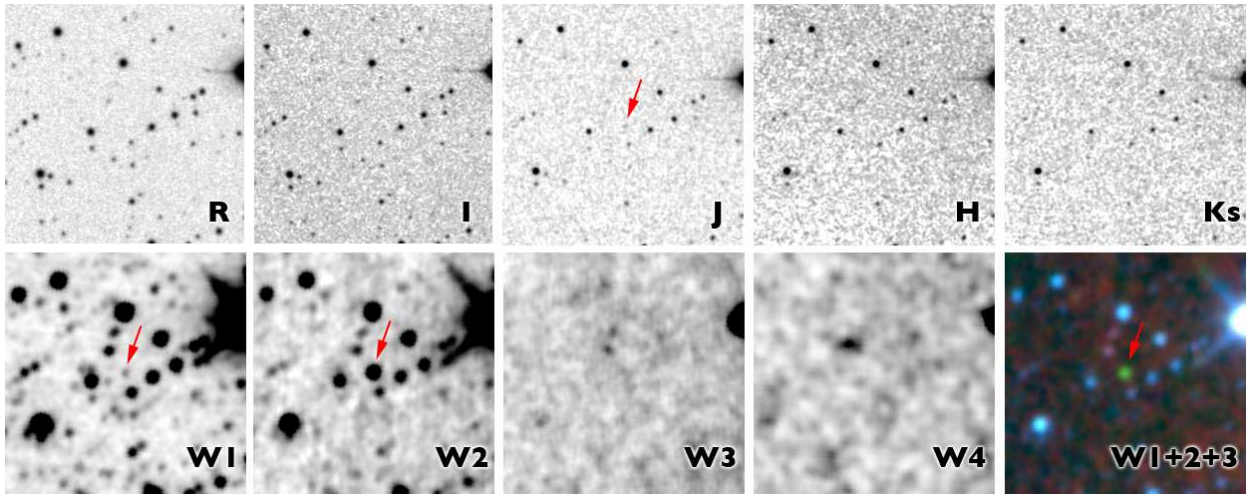}
\caption{\label{fig:0458_finder}Nine-band finder chart showing the area around the brown dwarf.  The top row shows DSS images at $R$ and $I$ bands taken at epochs 1989-12-23, and 1996-10-02 (UT, respectively), as well as 2MASS $J$, $H$ and $K$ bands taken at epoch 1999-01-01 (UT). The bottom row shows the WISE coadded images at $W1$, $W2$, $W3$, and $W4$, with an RGB composite of bands $W1$, $W2$ and $W3$.  The coadds come from frames ranging between MJDs 55261.554191 and 55262.811271.  Each image is 5 arcminutes on a side with north up and east to the left.  The brown dwarf is detected in $J$, $W1$, and $W2$ and shows significant motion between epochs; the $W3$ flux is contaminated by nebulosity and is therefore unreliable, and it is undetected in $W4$.  The final image is a three-color composite comprised of $W1$, $W2$ and $W3$, shown as blue, green and red, respectively.}
\end{figure}

\begin{figure}
\figurenum{4}
\plotone{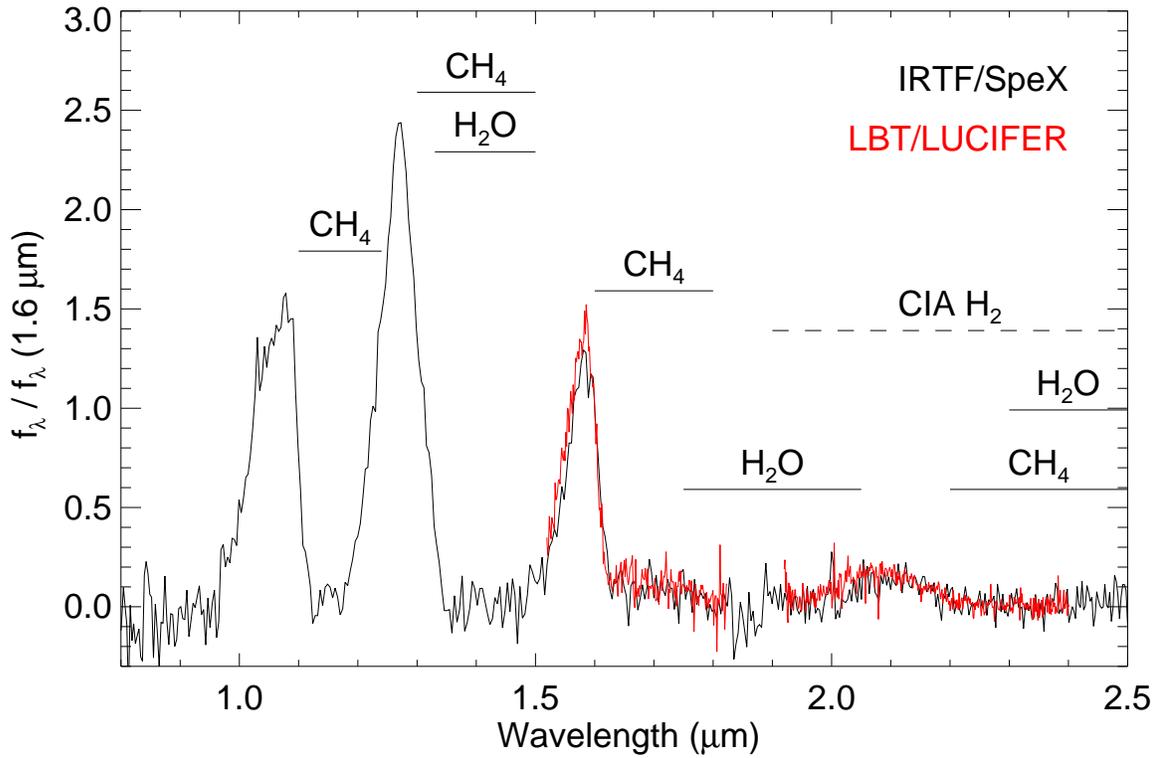}
\caption{\label{fig:spex_lucifer}Spectrum of \obj\ obtained with SpeX (\textit{black}) and LUCIFER (\textit{red}).  Prominent absorption bands of CH$_{4}$, H$_{2}$O, and H$_{2}$ are indicated.  The agreement between the two spectra is excellent.}
\end{figure}

\begin{figure}
\figurenum{5}
\plotone{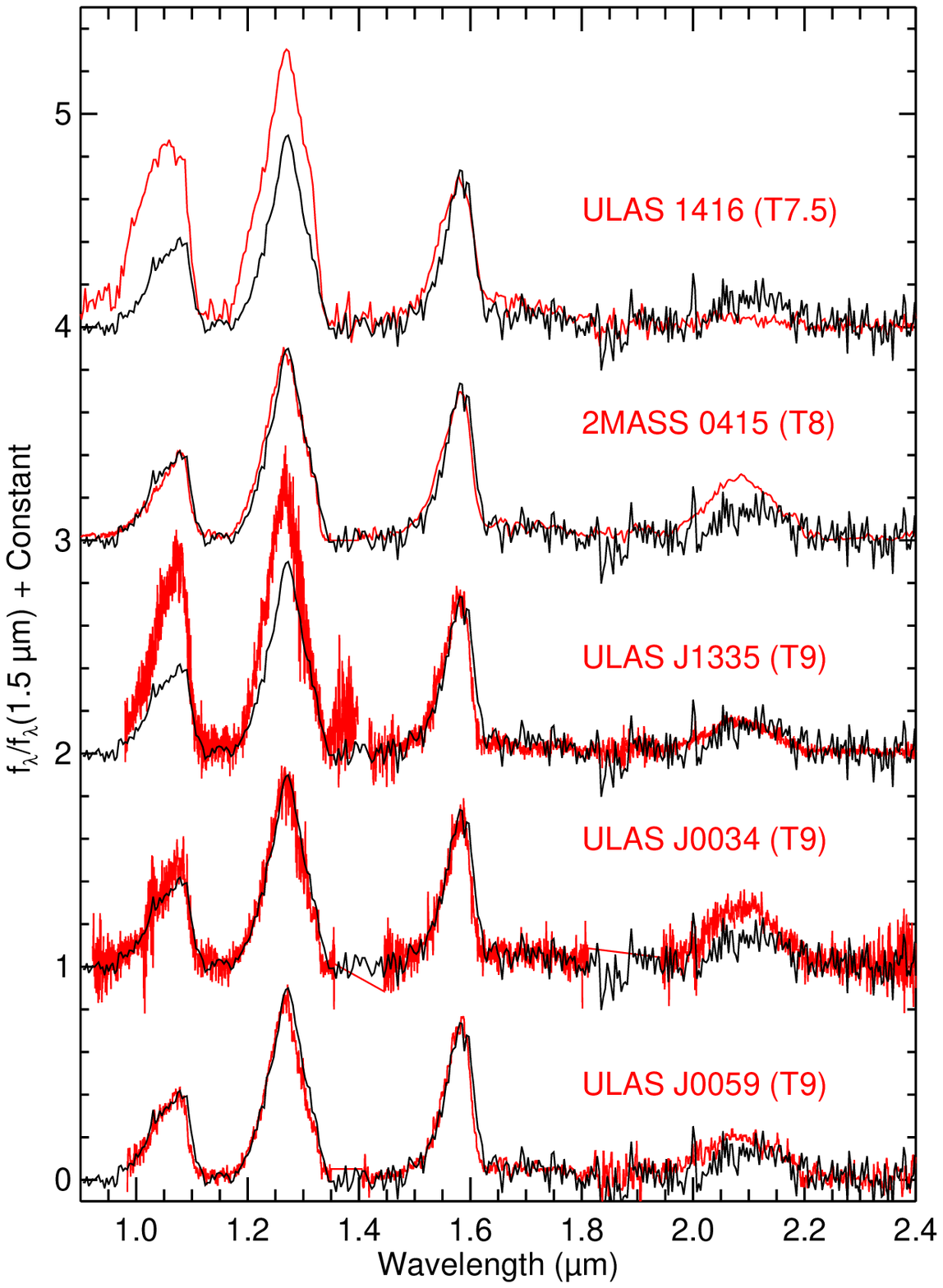}
\caption{\label{fig:spectral_comparison}The 1.5 $-$ 2.45 $\mu$m spectrum (black) of \obj\ .  Overplotted (red) are five late T dwarf spectra: ULAS 1416$+$13 \citep{Scholz,Burgasser10}, 2MASS J0415$-$09 \citep{Burgasser02}, ULAS J1335$+$11 \citep{Burningham08}, ULAS J0034$-$00 \citep{Warren} and CFBDS J0059$-$01 \citep{Delorme}.}
\end{figure}

\begin{figure}
\figurenum{6}
\plotone{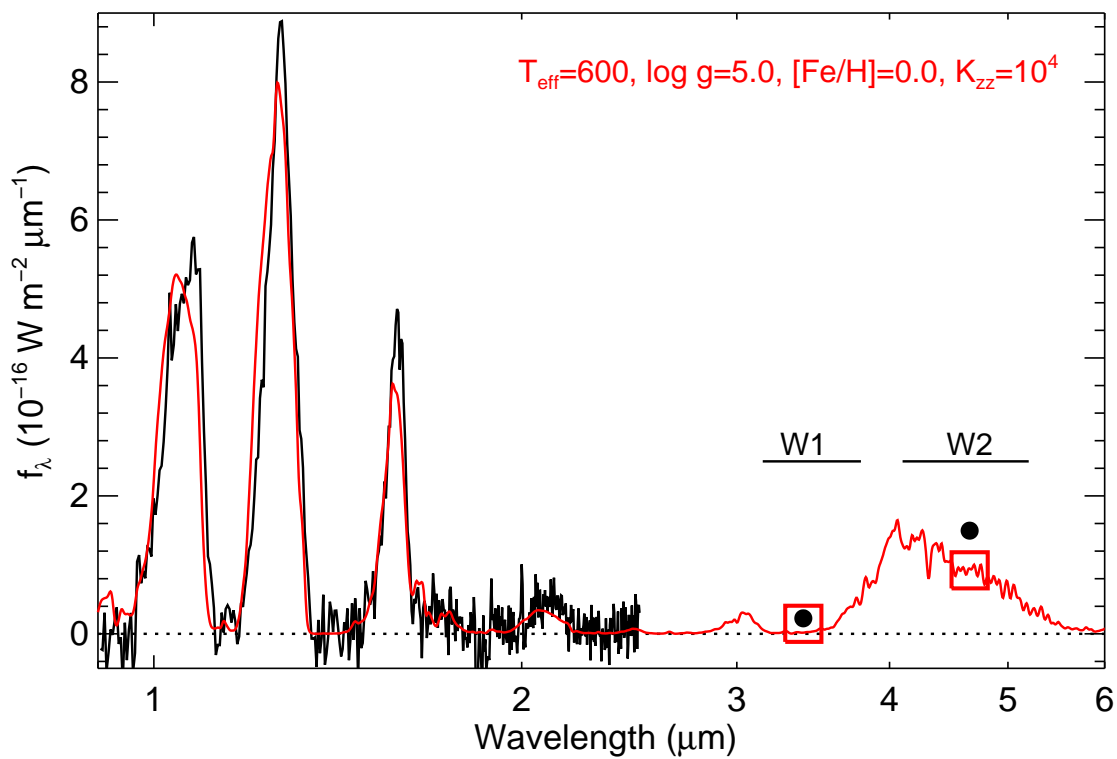}
\caption{\label{fig:ModComp}Spectrum and WISE photometry
  (\textit{black}) of \obj\ overplotted with a model spectrum.  The
  best fitting model spectrum for the case $w_i$=1 (red) has \teff=600 K,
  \logg=4.5, [Fe/H] = 0, and \kzz=10$^4$ cm$^2$ s$^{-1}$.  The average model
  fluxes over the WISE bandpasses are shown as open squares; the photometric errors on the WISE points are comparable to the symbol size.}
\end{figure}

\end{document}